\documentclass{llncs}
\usepackage{graphicx}
\usepackage{amsmath}
\usepackage{color}
\usepackage{balance}
\usepackage[T1]{fontenc}
\usepackage[ruled,vlined,linesnumbered]{algorithm2e}
\usepackage{oz}
\usepackage{url}
\usepackage{pifont}
\begin{document}

\title{Contextualisation of Data Flow Diagrams for security analysis}

\author{Shamal Faily \inst{1}\orcidID{0000-0002-2859-1143} \and Riccardo Scandariato \inst{2} \and Adam Shostack \inst{3} \and Laurens Sion \inst{4}\orcidID{0000-0002-8126-4491}  
     \and Duncan Ki-Aries \inst{1}\orcidID{0000-0001-8114-2737}}
\authorrunning{Faily et al.}
\institute{Department of Computing \& Informatics, Bournemouth University, Poole, UK\\
\email{\{sfaily,dkiaries\}@bournemouth.ac.uk}
\and
Chalmers and University of Gothenburg, Gothenburg, Sweden\\
\email{riccardo.scandariato@cse.gu.se}
\and
Shostack \& Associates, Seattle, USA\\
\email{adam@shostack.org}
\and
imec-DistriNet, KU Leuven, Leuven, Belgium\\
\email{laurens.sion@cs.kuleuven.be}
}

\maketitle

\begin{abstract}
Data flow diagrams (DFDs) are popular for sketching systems for subsequent threat modelling.  Their limited semantics make reasoning about them difficult, but enriching them endangers their simplicity and subsequent ease of take up.  We present an approach for reasoning about tainted data flows in design-level DFDs by putting them in context with other complementary usability and requirements models.  We illustrate our approach using a pilot study, where tainted data flows were identified without any augmentations to either the DFD or its complementary models.  
\end{abstract}

\section{Introduction}
\label{sect:intro}
Data Flow Diagrams (DFDs) are useful as a sketch that explores how a system and its elements might be exploited; their simplicity makes it possible for different people with different levels of expertise to contribute to the security analysis of a system as it is evolves.  

As DFDs become more critical to security design practices, so too is the need to reason about their properties using software tools. 
Limitations around cognitive ability, expertise and time constrain the effectiveness of modellers when scaling up or making decisions around DFDs~\cite{sim79}.  However, their limited semantics makes reasoning with DFDs alone difficult; this leads to an inherent trade-off between using easy to adopt notations and those that afford automated reasoning but are more elaborate~\cite{syvv20}.

Data flows are analogous with information flows.  Information flow analysis (like taint analysis) is a long established technique for reasoning about the interactions of data within entities, and their impact on security as the data flows through the system \cite{denn79,yiso07}.  Unfortunately, visual inspection alone is insufficient for spotting potential issues with data inside data flows.  Formal policy specifications and binary instructions provide the context necessary to reason about tainted information flows, but DFDs lack this level of precision. The options are either (i) adding additional information to the diagram itself, or (ii) providing context via other models aligned with DFDs.
In the related work, the first route has been extensively explored~\cite{tcs18,tusb19}, so this paper takes the less followed second path.
\emph{Usability models} could play a particularly important role in providing such context. For example in \cite{fail18}, usability models describe the main tasks performed by a software system, and the roles associated to those tasks. The models relate to the overall goals and requirements of the system. Just as DFDs provide early insights into how systems might be exploited, usability models indicate where interaction problems might subsequently facilitate exploitation.  These different models might be produced independently and, with inter-operable tools, we can reason about the security impact these models have on DFDs, and vice-versa.  

\textbf{Contribution}. In this short paper, we present an approach for identifying potential taint in design-level DFDs.  Our guiding principle is that, to encourage adoption, DFDs should be no more graphically complex than they currently are.  Instead, we should leverage the alignment between DFDs and other usability and requirements models.  We present the related work upon which our approach is based in Section \ref{sect:background} before presenting the key concepts and algorithms in our approach in Section \ref{sect:approach}.  We illustrate our approach in Section \ref{sect:casestudy} by using it to identify pre-process and post-process taint in a critical infrastructure pilot study, before discussing the implications of this work in Section \ref{sect:discussion}.

\section{Related Work and Background}
\label{sect:background}

\subsection{Reasoning about Data Flow Diagrams in Threat Modelling}

Data Flow Diagrams (DFDs) graphically model flows of information (data flows) between human or system actors external to a system (entities), activities that manipulate data (processes), and persistent data storage (data stores) \cite{yoco79}.  This notation is often extended with trust boundaries: dotted boxes encompassing DFD elements operating at the same level of privilege.  Trust boundaries help identify data flows that cross privilege levels \cite{shos14}.

DFDs have overlapping functions.  Diane (a diagram creator) creates a DFD that diagrammatically represents her mental model.  On viewing the DFD, Elaine (an engineer) internalises this mental model and requests changes.  Dialogue around their differences subsequently brings both mental models closer together.   
Francis (a formal modeller) crafts a structured representation of a system, from which subsequent reasoning can be performed.  This relationship between a mental model, a diagram, and a formal model has not been well explored.

Tuma et al. \cite{tsws18} first examined the potential of using information flow analysis to reason about DFDs.  They extended the DFD notation by labelling data flows with assets and their security properties, indicating the source and target of assets, including domain properties and assumptions from the KAOS modelling language \cite{lams09}.  In later work, Tuma et al. \cite{tusb19} further illustrate the potential for using DFDs for design-level information flow analysis.  In their approach, a domain specific language is used to model DFDs annotated with security labels.  The model is subsequently rendered as a graph and statically analysed.

Antigac et al. \cite{anss18} examined how certain properties of a DFD can be hotspots for further investigation.  For example, a usage hotspot corresponds with 3 DFD elements: data flow $d$ into process $p$, process $p$, and data flow $d'$ from $p$.  Antigac et al. showed how such hotspots bridge the gap between different models, and provide a basis for subsequent model transformation without fundamentally changing the visual semantics of DFDs.

\subsection{Security and Software Design Meta-models}

\begin{figure}[h!]
\centering
\includegraphics[scale=0.45]{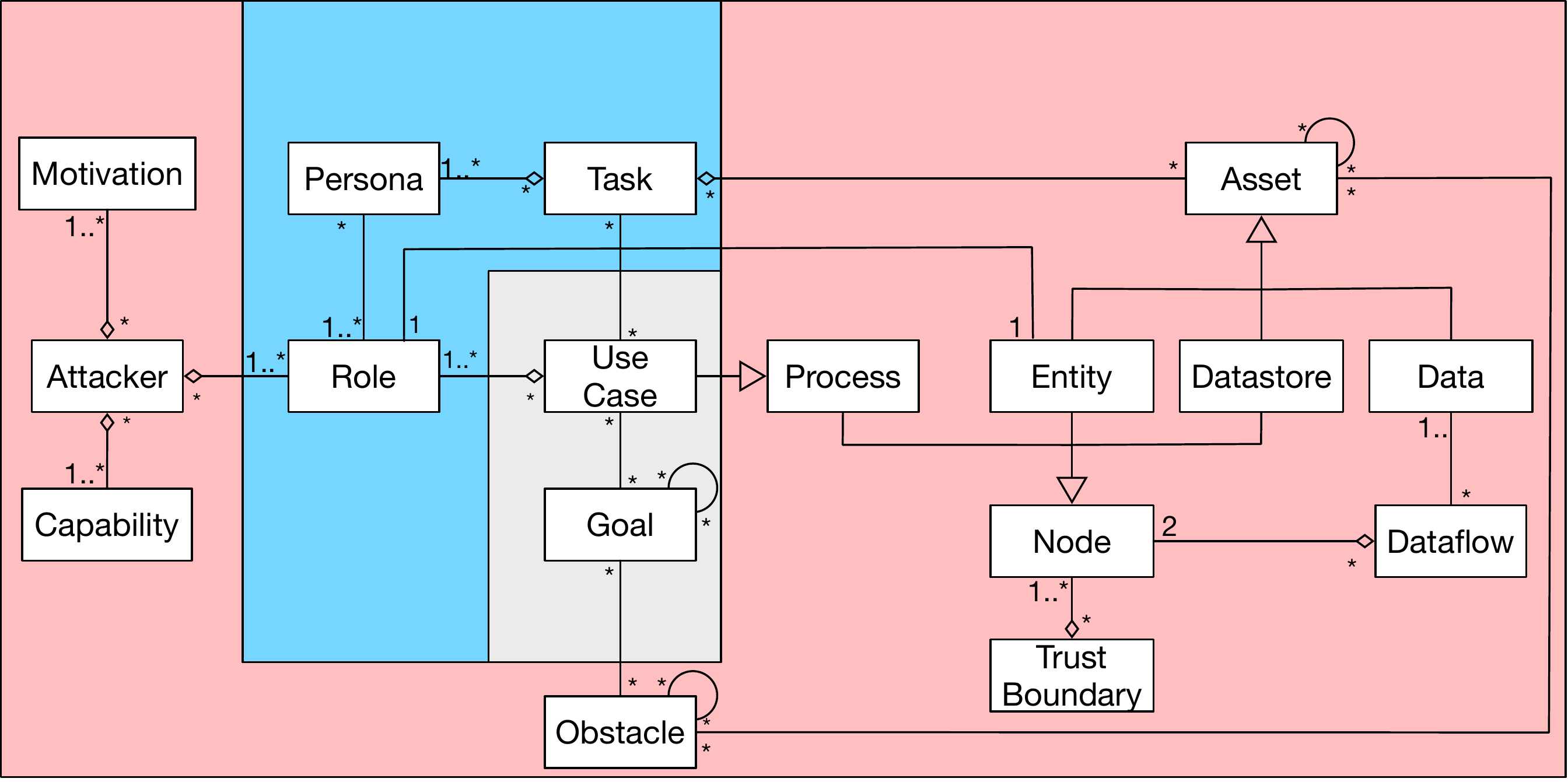}
\caption{A UML class diagram showing the IRIS concepts related to threat modelling (red), usability (blue) and requirements modelling (grey)}
\label{fig:summary}
\end{figure}

Meta-models specify how model concepts are associated.  In doing so, they guide analysts in collecting and analysing model data, and guide tool builders in constructing tools to support them.  The software engineering community has examined the relationship between software and requirement modelling approaches and security, as summarised by \cite{matu17}.  These approaches do not, however, account for the role played by usability data and models.  The \textbf{IRIS (Integrating Requirements and Information Security)} meta-model was devised to provide guidance on how early-stage design concepts from usability as well as security and requirements engineering might be aligned \cite{fail18}. A sub-set of the IRIS concepts relevant to this paper is provided in Figure \ref{fig:summary}.

Coles et al.~\cite{cofk18} demonstrated how use cases and assets provide the concepts necessary to threat model with data flow diagrams, and how -- in addition to modelling system goals -- the KAOS modelling language~\cite{lams09} is also suitable for modelling attack trees as obstacles.  To make attacker assumptions more explicit, IRIS supports the specification of attackers.  Attackers need not be intrinsically malicious, but they will have some motivations as drivers for carrying out an attack, and capabilities that provide the knowledge and resources necessary to mount and sustain any threat.  IRIS draws its taxonomy of motivations from \cite{van07}, and capabilities from \cite{joas05}.  An additional motivation of productivity was also added to better reflect non-malicious attackers who intentionally or unintentionally commit harm to get their job done.   

To leverage the outputs of user research in security design, two popular usability modelling concepts are supported by IRIS.  Personas are specifications of archetypical user behaviour \cite{core14}; they not only capture user goals and expectations, but their construction and usage helps elicit security requirement \cite{fafl106}.  Tasks are narrative scenarios that describe both the personas and the broader system -- including use cases -- in context. 

\section{Approach}
\label{sect:approach}

Our approach focuses on how \emph{tainted} data flows cast doubt on the safety of the data they carry.  Unlike traditional taint analysis on program source code, the origins of data flow taint in our approach could be human error resulting from human entities and processes, or issues resulting from the DFDs and associated specifications.  These problems could have an indeterminate impact on affected endpoints, thereby warranting further investigation.  Aligning DFDs with usability and requirements models provides context to assist such an investigation.

Assuming the pre-requisite models exist, our approach validates them using the analysis checks described in Section \ref{sect:checks}.  Because of its alignment with the DFD concepts as shown in Figure \ref{fig:summary}, our approach relies on the IRIS meta-model.  DFD processes are analogous with use cases, and actors in use cases could be human or system entities.  DFDs directly link to usability models because use cases, as processes, put tasks in context.  DFDs are also indirectly linked because roles constituting use case actors are also fulfilled by personas -- who interact in tasks -- putting these roles in context.

\subsection{Dataflow specification}

DFDs are graphs, but can be specified as a set of data flow types.  In our approach, a data flow consists of a label, names of the DFD elements data flows from and to, and the types of these elements, where $NODE$ is either an $entity$, a $process$, or a $datastore$.  Data flows also specify the information assets (as $DATA$) they carry.  Using Z \cite{woda96}, we can express a data flow formally, where the predicate part of the schema contains the well-formedness constraints:

\scriptsize
\begin{schema}{DataFlow}
  label, from, to: STRING \\
  fromType, toType: NODE \\
  assets: \pset DATA
\ST
  assets \neq \emptyset \\
  ((fromType = entity) \land (toType = process)) \lor\\
  ((fromType = process) \land (toType = entity)) \lor\\   
  ((fromType = datastore) \land (toType = process)) \lor\\      
  ((fromType = process) \land (toType = datastore)) \lor\\ 
  ((fromType = process) \land (toType = process))     
\end{schema}
\normalsize

\subsection{Pre-Process and Post-Process analysis}
\label{sect:checks}
For each entity in the DFD, our approach first visits the entity's data flows using the $dataFlows$ recursive graph traversal function described in Algorithm~\ref{alg:dfs}.  The function populates a persistent array of unique data flow sequences ($allSeqs$), and a persistent set of previously visited DFD elements ($visited$). 

\begin{algorithm}
  \SetKwInOut{Input}{Input}
  \scriptsize
  \SetAlgoLined
  \Input{currentNode - $NODE$, prefix - $\seq DataFlow$}
  \KwData{allSeqs - $\seq (\seq DataFlow)$, visited - $\power NODE$, nodeFlows - $\ran Node \rel DataFlow$}
  
  \SetKwProg{Fn}{Function}{ is}{end}
  \Fn{dataFlows($currentNode$, $prefix$)}{
  
  $visited$.add($currentNode$);
  
  $dfs$ $\leftarrow$ nodeFlows\ $currentNode$;
  
  \eIf{$dfs$ = $\emptyset$} {  
    \If{prefix.length > 0}{
       $allSeqs$.append($prefix$);        
    }    
  }
  {    
    \While{$df$ $\leftarrow$ $dfs$} {
      $newPrefix$ $\leftarrow$ $prefix$;
      
      $newPrefix$.append($df$);
      
      \eIf{$df$.to $\in$ $visited$}{
          $allSeqs$.append($newPrefix$);
      }
      {
        dataFlows\ $df$.to\ $newPrefix$;
      }   
    }     
  }
  \textbf{return};
}
\caption{Identification of data flows}
\label{alg:dfs}
\end{algorithm}

\begin{algorithm}
  \SetKwInOut{Input}{Input}
  \scriptsize
  \SetAlgoLined
  \Input{dfSeq - $\seq DataFlow$}
  \KwData{contextualisedTask - $\ran\ UseCase \rel Task$, taskAsset - $\ran\ Task \rel Asset$, personaRoles - $\ran\ Persona \rel Role$, taskPersonas - $\ran\ Task \rel Persona$, roleAttackers - $\ran\ Role \rel Attacker$, allAttackerRoles - $\ran\ roleAttackers\inv$, attackerMotivation - $\ran Attacker \rel Motivation$, attackerCapability - $\ran Attacker \rel Capability$, taskDemand - $\ran Task \rel Value$, goalConflict - $\ran Task \rel Value$, processExceptions - $\ran UseCase \rel Obstacle$, obstructedGoals - $\ran Obstacle \rel Goal$, obstacleAssets - $\ran Obstacle \rel Asset$, nameToProcess - $String \pfun UseCase$, logPreProcessTaint - logs taint to process resulting from named task, logPostProcessTaint - logs taint to process resulting from named obstructed goal}
  
  \SetKwProg{Fn}{Function}{ is}{end}
  \Fn{analyseDataFlows($dfSeq$)}{
    \While{$df$ $\leftarrow$ $dfSeq$} {
    
      \tcc{Check for pre-process taint}
      \If{$df$.fromType = entity $\land$ $df$.toType = process $\land$ $df$.fromName $\in$ $Role$}{ 
                
        \While{$t$ $\leftarrow$ contextualisedTask\ (nameToProcess\ $df$.toName)} {     
          \If{$df$.assets $\cap$ taskAssets\ $t$} {
            
            \While{$r$ $\leftarrow$ (personaRoles\ (taskPersonas\ $t$) $\cap$ allAttackerRoles)} {
              
                \While{$a$ $\leftarrow$ roleAttackers\ $r$} {

                   \If{(Productivity $\in$ attackerMotivation\ $a$) $\land$ (Low\ Time  $\in$ attackerCapability\ $a$) $\land$ ( (taskDemand\ $t$ $\cap$ \{Medium,High\}) $\lor$ (goalConflict\ $t$ $\cap$ \{Medium,High\}) ) }{

                    logPreProcessTaint\ (nameToProcess\ $df$.toName)\ $t$;
               
                  }                  
                }     
            }  
          }
        }
      }        
         
       \tcc{Check for post-process taint}
       \If{$df$.fromType = process} {
         \While{$o$ $\leftarrow$ processExceptions\ $df$.fromName} {
  
           \If{(obstacleAssets $o$ $\cap$ $df$.$assets$) $\neq$ $\emptyset$}{
             \While{$g$ $\leftarrow$ obstructedGoals\ $o$} {

                \If{isObstacleObstructed\ $o$ = true } {
                  logPostProcessTaint\ (nameToProcess\ $df$.fromName)\ $g$;
                }
    
             }
           }
         }
       }
     }
     \textbf{return};
  }
\caption{Taint analysis}
\label{alg:adf}
\end{algorithm}

Each sequence in $allSeqs$ is then enumerated to identify and log potential data \emph{pre-process} and \emph{post-process} taint as described in Algorithm \ref{alg:adf}. The types mentioned in the algorithm can be found in Figure \ref{fig:summary}, with the exception of $VALUE$, where $VALUE \ddef Low \bbar Medium \bbar High$.

Pre-process taint checks (lines 3--15) identify instances where means, motives, and opportunity are present for human errors and violations.  The checks are performed on data flows going from human entities to processes contextualised as tasks; these processes are use cases linked to tasks as indicated in Figure  \ref{fig:summary}.  Tasks become a possible source of human error when three conditions hold.  First, roles fulfilled by personas in a task are shared with roles fulfilled by attackers.  Second, attackers have a non-malicious motive and are constrained in the means available; we define such attackers as motivated by productivity and, as a capability, a limited amount of time.  Finally, affected tasks are demanding to the affected personas, or in tension with their personal goals.

Post-process taint checks (lines 16--26) identify instances where exceptions resulting from processes are unresolved, and these exceptions impact information flowing from processes.  Exceptions are modelled as obstacles obstructing one or more system goals operationalised as the affected processes.  An obstacle impacts an out-going data flow if assets associated with the obstacle intersect with information assets in the data flow.  An exception is unresolved if these obstacles are not resolved by another goal, as determined by the $isObstacleObstructed$ function defined in Algorithm \ref{alg:ioo}.  It begins by determining whether the input obstacle has been resolved by another goal.  After evaluating whether the obstacle has been resolved, the check enumerates both obstacles that are or-refined and and-refined.  In the case of or-refined obstacles, an obstruction on \emph{any} of the refined obstacles is enough to consider the obstacle obstructed.  Conversely, in the case of and-refined obstacles, an obstruction is present only if \emph{all} refined obstacles are obstructed.  
  
\begin{algorithm}
  \scriptsize
  \SetKwInOut{Input}{Input}
  \SetKwInOut{Output}{Output}
  \KwData{resolvedObstacles - $\ran Obstacle \rel Goal$, orRefinedObstacles - $\ran Obstacle \rel Obstacle$, andRefinedObstacles - $\ran Obstacle \rel Obstacle$}

  \SetAlgoLined
  
  \Input{o - the obstacle name}
  \Output{isObstructed - indicates if obstacle $o$ is obstructed}

  \SetKwProg{Fn}{Function}{ is}{end}
  \Fn{isObstacleObstructed($o$)}{
  
    $ros$ $\leftarrow$ resolvedObstacles\ $o$;
		
    \eIf{$ros$ $\neq$ $\emptyset$}{
      isObstructed $\leftarrow$ false; 			
    }
    {
      $obs$ $\leftarrow$ orRefinedObstacles\ $o$;
			
      \While{$oro$ $\leftarrow$ $obs$}{
        $isObstructed$ $\leftarrow$ isObstacleObstructed\ $oro$;

        \If{$isObstructed$ = true}{
          break;
        }
      }
			
	  $obs$ $\leftarrow$ andRefinedObstacles\ $o$;

	  \While{$aro$ $\leftarrow$ $obs$}{
        $isObstructed$ $\leftarrow$ isObstacleObstructed\ $aro$;
				
	    \If{$isObstructed$ = false}{
	      break;        
	    }        
      }
    }
    \textbf{return} $isObstructed$;
  }
\caption{isObstacleObstructed check}
\label{alg:ioo}
\end{algorithm}

\subsection{Implementation}

We have demonstrated the feasibility of our approach by implementing it in CAIRIS release 2.3.3.  CAIRIS (Computer-Aided Integration of Requirements and Information Security) is an open-source software platform for eliciting, specifying and validating secure and usable system specifications \cite{cairis} developed as an exemplar for IRIS tool-support.
 
 CAIRIS models, once imported into the platform, are implemented as relational databases.  Graphical models in CAIRIS are automatically generated using a pipeline process, where a declarative model of graph edges is generated by CAIRIS; this is processed and annotated by graphviz \cite{graphviz} before being subsequently rendered as SVG.  SQL stored procedures implement a suite of security and privacy model validation checks.  Algorithms 1 - 3 were implemented as SQL stored procedures; these are executed during a normal model-validation check.  No changes were made to pre-existing visual models and the IRIS meta-model.

\section{Pilot Study: Modifying telemetry outstation software}
\label{sect:casestudy}

We used our approach to identify process taint in a partial specification of a software repository for industrial control software.  While based on a hypothetical water treatment company, this anonymised specification is drawn from a more complete specification model created for a UK water treatment company.  The CAIRIS model\footnote{Available from \url{https://doi.org/10.5281/zenodo.3872071}} of this partial specification consists of 1 attacker, 1 role, 1 persona, 1 task, 1 use case, 28 goals, 17 obstacles, 58 goal and obstacle associations, 11 assets, 11 asset associations, and 7 data flows.  Creation of the model is not the subject of this paper, but further details of how the broader model was created are provided in \cite{fafl103}.  

The specification captures the system goals and complementary model elements associated with modifying software running on telemetry outstations.  Such outstations provide the means for remotely monitoring and controlling physical infrastructure such as water pumps.  Malicious tampering of such outstations contributed to the well publicised Maroochy Water Breach \cite{slmi07}.   
\begin{table}
\parbox{.4\linewidth}{
\scriptsize
\center
\begin{tabular}{|p{3cm}|p{3cm}|}
\hline
\textbf{Dataflow} & \textbf{Assets}                                                                                           \\ \hline 
job    & Job\\ \hline
software (to Sandbox)  &  Telemetry Software File\\ \hline
software (from Sandbox)  &  Telemetry Software File\\ \hline
updated software  &  Telemetry Software File\\ \hline
current software  &  Telemetry Software File\\ \hline
alarm &  Alarm\\ \hline
update &  Software Change\\ \hline
\end{tabular}
\caption{Dataflows and assets}
\label{tab:dfa}
}
\hfill
\parbox{.55\linewidth}{
\scriptsize
\center
\begin{tabular}{|p{0.5cm}|p{2.1cm}|p{1cm}|p{1cm}|}
\hline
\textbf{Id} & \textbf{Sequence} & \textbf{Pre-Proc.} & \textbf{Post-Proc.}                                                                                             \\ \hline 
1 & $\langle$job, alarm$\rangle$ & \ding{55} & \ding{55}  \\ \hline
2 & $\langle$job, update$\rangle$ &  \ding{55}  & \checkmark \\ \hline
3 & $\langle$job,updated software,current software$\rangle$ & \ding{55}  & \checkmark \\ \hline
4 & $\langle$job,software, software$\rangle$ &  \ding{55}  & \checkmark  \\ \hline
5 & $\langle$current software$\rangle$& \checkmark  & \checkmark  \\ \hline
\end{tabular}

\caption{Dataflow sequences and results of pre-process and post-process taint checks}
\label{tab:seqs}
}
\end{table}

\begin{figure}[h!]
\centering
\includegraphics[scale=0.6]{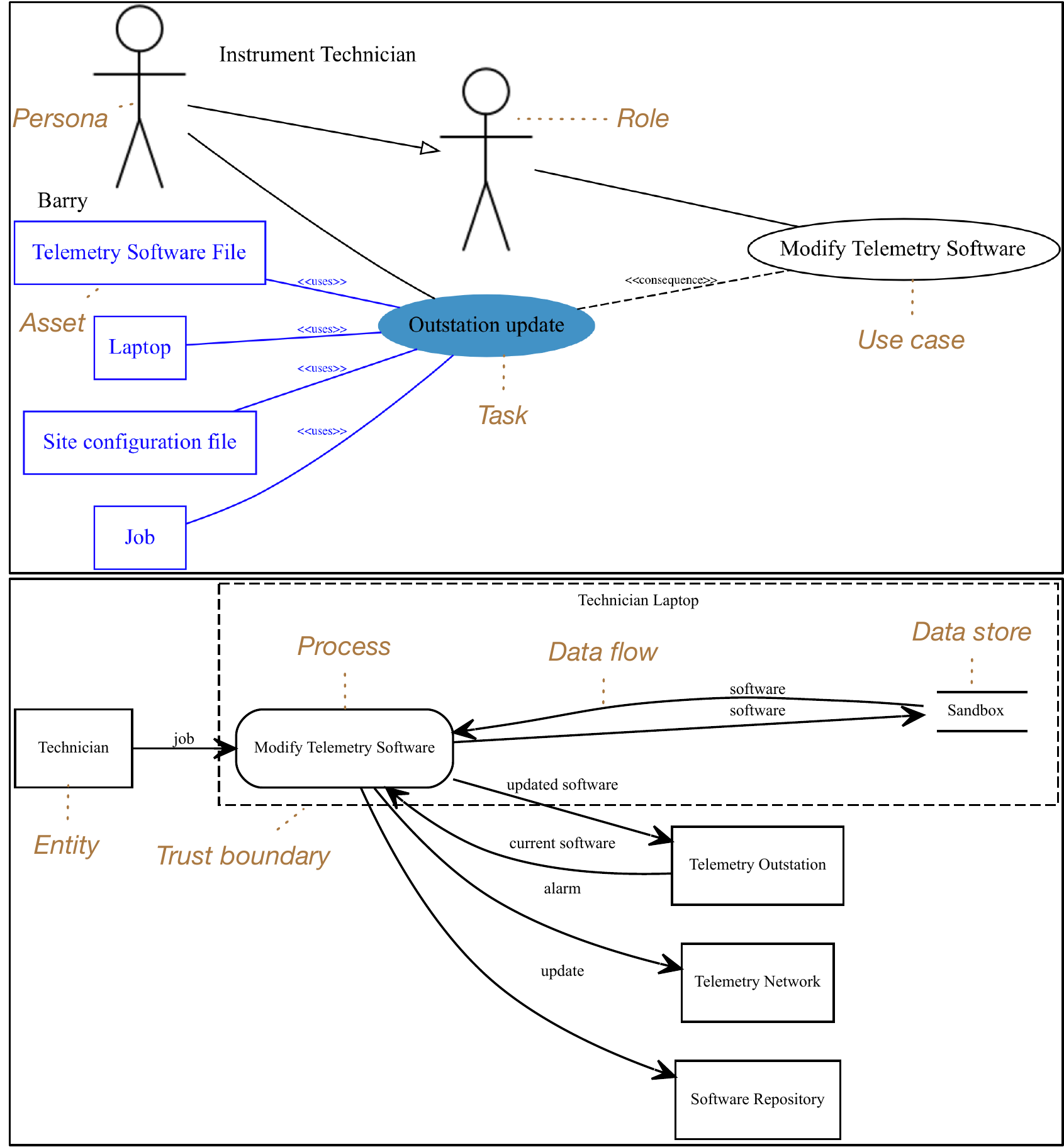}
\caption{Usability model (top) and DFD (bottom) of Modify Telemetry Software generated by CAIRIS}
\label{fig:dfd}
\end{figure}

Our pilot study considers the impact of human error by an overworked technician focusing on the intricate task of updating software on telemetry outstations (Outstation update). This task puts in context the use case Modify Telemetry Software as shown in Figure \ref{fig:dfd} (top), which is carried out by an instrument technician persona (Barry).  Details of how the persona and tasks were constructed are described in more detail in \cite{fafl106}.  The task model provided the context necessary to model the DFD generated by CAIRIS in Figure \ref{fig:dfd} (bottom).  Table \ref{tab:dfa} specifies the assets carried in each data flow.   

\begin{figure}[h!]
\centering
\includegraphics[width=\textwidth]{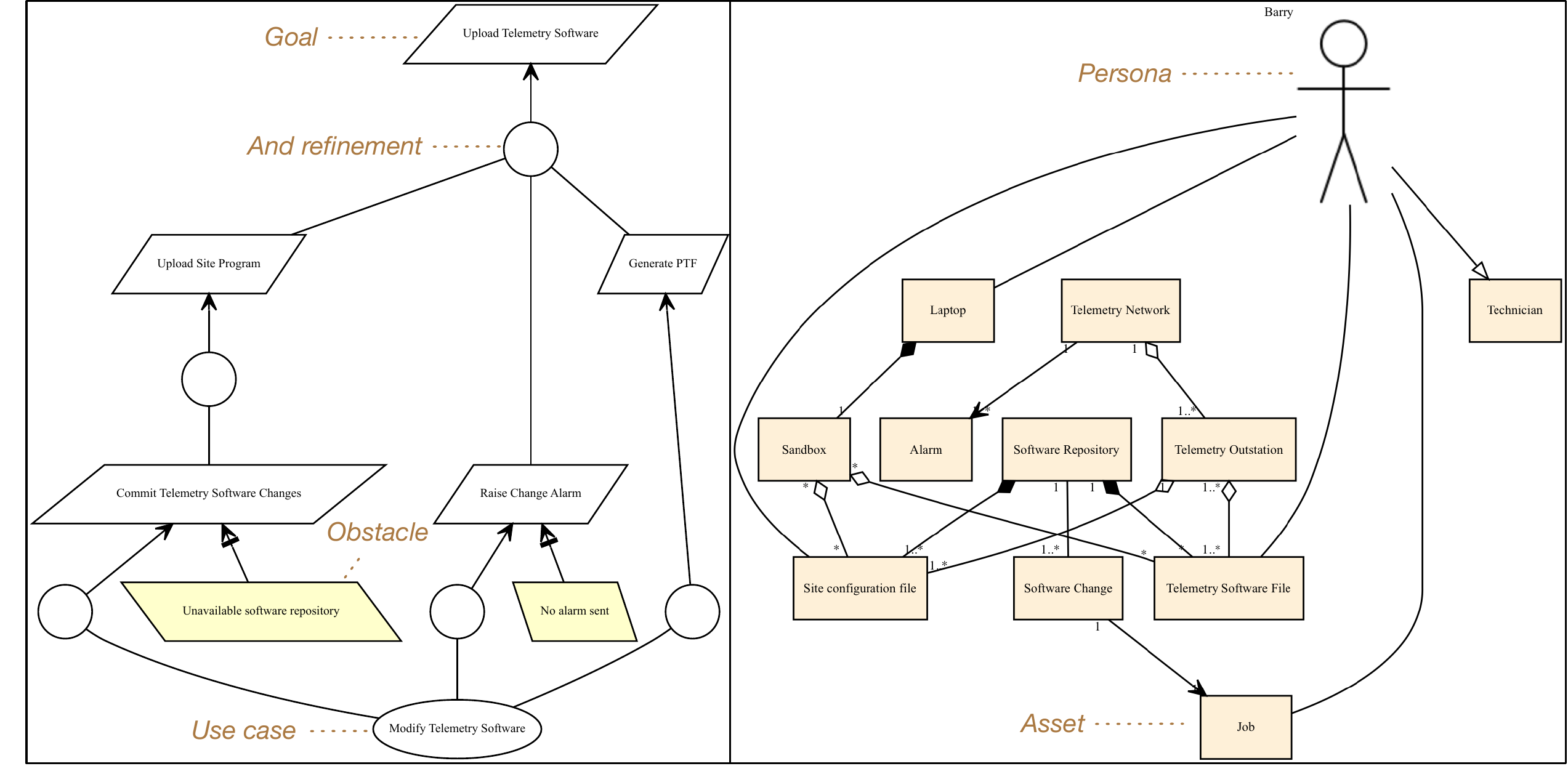}
\caption{Complementary KAOS goal model (left) and UML class diagram-based asset (right) model generated by CAIRIS}
\label{fig:ga}
\end{figure}

Not shown in the visible models is an attacker (Unintentional Barry).  This attacker's motivation and resources are specified as `Productivity' and `Low Resources/Personnel and Time' to reflect non-malicious intent and a busy schedule.  The task model also indicates the assets that Barry directly or indirectly interacts with in completing this task.  The relationship between these and other assets associated with the specification are shown in Figure \ref{fig:ga} (right).

On performing a model validation check, five unique sequences of data flows were generated as shown in Table \ref{tab:seqs}.  The check indicates pre-process taint associated with sequences 1, 2, 3, and 4 resulting from the flow between the technician and the process.  This was due to the $job$ flow carrying alarm information associated with the task and the potential for error.  The task narrative describes how Barry needs to raise an alarm to validate the setup is correct; the alert draws attention to the implications of not safeguarding this information asset.

The model validation check also indicates post-process taint associated with Sequence 1; this outgoing process flow carries alarm information.  An exception is associated with the second step of the process, where the system sends a change alarm.  As a cut of the goal model in Figure \ref{fig:ga} (left) shows, the associated obstacle remains unresolved and, although not visible, the obstacle is concerned with the alarm asset carried in $job$.

\section{Discussion and Conclusion}
\label{sect:discussion}

This short paper showed how, by putting DFDs in context, we can identify process taint without changing any DFD semantics.  CAIRIS demonstrates the feasibility of our approach, but it could be adapted to any inter-operable combination of tools.  Solutions for resolving the problems are not prescribed besides changing the attacker model and tasks, or resolving exceptions.  However, by indicating otherwise invisible problems, our approach sheds light on why problems exists, and how a system or its context of use might need to change to address them.  This approach is contingent on specifications containing the concepts in Figure \ref{fig:summary} that might be created before, during, or after DFD creation.  Small or poorly resourced teams may lack the resources to maintain such models given the user research investment required.  However, this approach does allow human factor experts to become more engaged with threat modelling.  We are currently working with system engineering teams with such expertise to evaluate the impact this approach has on increasing such engagement. 

A threat to validity is the small size of the pilot study specification.  However, we have also evaluated our approach using a more complex military medical evaluation system model described in \cite{kfdw18} consisting of 10 attackers, 14 roles, 9 personas, 12 tasks, 29 use cases, 46 goals, 25 obstacles, 167 goal and obstacle associations, 82 assets, 388 asset associations, and 134 data flows.  No differences in model validation performance were noted for this larger model, but a detailed evaluation of this and other larger models will be the subject of future work. 

Our approach only considers non-malicious attackers engaging in difficult tasks.  However, Algorithm \ref{alg:adf} can be extended to consider alternative attacker and task attributes corresponding with different means, motives, and opportunities.  For example, an inside attacker might be motivated by improved esteem or thrill seeking, and participate in tasks with differing levels of goal conflict. 

\subsubsection*{Acknowledgements}

This paper resulted from discussions at Dagstuhl Seminar 19231: Empirical Evaluation of Secure Development Processes. 

\bibliographystyle{splncs04}
\bibliography{gramsec20}
\end{document}